\documentclass[a4paper, twocolumn, nofootinbib, showpacs, amsmath, amssymb]{revtex4}

\usepackage{graphicx}

\begin{document}

\title{ Implications of minimum-length deformed quantum mechanics for QFT/QG \footnote{Extended version of the talk delivered at the workshop dedicated to S.~G.~Matinyan's 80th birthday, 28-29 September, 2011, Tbilisi \\ \textsuperscript{$\star$}Email:~~maziashvili@gmail.com  \\ \textsuperscript{$\dagger$}Webpage:~~www.cepp.iliauni.edu.ge }}

\author{ Michael~Maziashvili \textsuperscript{$\star$} }

\affiliation{ Center for Elementary Particle Physics, ITP, Ilia State University, 3/5 Cholokashvili Ave., Tbilisi 0162, Georgia \textsuperscript{$\dagger$} }

\begin{abstract}

After picking out what may seem more realistic minimal gravitational deformation of quantum mechanics, we study its back reaction on gravity. The large distance behaviour of Newtonian potential coincides with the result obtained by using of effective field theory approach to general relativity (the correction proves to be of repulsive nature). The short distance corrections result in Planck mass black hole remnants with zero temperature. The deformation of position-momentum uncertainty relations leads to the superluminal motion that can be avoided by making similar deformation of time-energy uncertainty relation. Such deformation also avoids UV divergences in QFT.

\end{abstract}

\pacs{04.60.Bc }


\maketitle

\section{Introduction}
\label{intro}

Nowadays, the ultimate theory of quantum gravity is still "under construction". Nevertheless, there are several different successful approaches to quantum gravity, each of these having certain advantages over the others for studying of a concrete question \cite{Rovelli:2000aw}. An unifying framework awaits its discovery. The present article mainly focusses on a phenomenological discussion of minimum-length deformed uncertainty relations. As the concept of minimum length naturally appears in all approaches to quantum gravity, one might hope that the bottom-up study by incorporating of minimum length into quantum mechanics may show up a good deal of knowledge about quantum gravity. Or otherwise speaking, the minimum length deformed quantum mechanics may prove to be a very economic and at the same time quite predictive unifying framework for discussing quantum gravitational effects.

It is evident that even in the simplest case there is a good deal of freedom in possible minimum-length deformations of position-momentum uncertainty relations(s). In the beginning we point out that one of them is naturally singled out and provide further motivation for it. Indeed, for this sort of modification there are several well known {\tt Gedankenexperiment} approaches by combining basic principles of quantum mechanics with general relativity. Then using a particular Hilbert space representation of such a deformed quantum mechanics, we address the question of consequent modification of QFT both at first and second quantization levels and subsequently its possible back reaction on gravity. (Throughout this discussion the gravitational interaction is treated as just produced by exchanging of graviton and the modifications due to minimum-length deformed quantum mechanics is assumed to be universal for all particles, including graviton. Shortly speaking, in estimating corrections to the Newtonian potential the linearized gravity with respect to the Minkowskian background is assumed, but then some of the results are extrapolated beyond this approximation as we do not yet know how to implement the deformed momentum operator that comes from minimum-length deformed quantum mechanics into the Einstein-Hilbert action). A curious result seems to be the existence of zero-temperature black hole remnants for such modified gravity as well as the recovery of well known quantum-gravitational black-hole entropy corrections. In the end we briefly address the question of UV finiteness of QFT as long as the time-energy uncertainty relation is also deformed and the effect of space-time dimension running/increase in view of the gravitationally deformed quantum mechanics.

\section{Implementing minimum length in quantum mechanics}
\label{implmlqm}

On dimensional grounds one can consider various gravitational corrections to the Heisenberg uncertainty relation that result in the lower bound on position uncertainty (we assume $c=1$, that is, $[\hbar]=$g$\cdot$cm,\,$[G_N]=$cm$/$g)

\begin{eqnarray}\label{gur1}\delta X \delta P \, &\geq & \, \frac{\hbar}{2} \,+\, \sum\limits_i\beta_i\hbar^{(\alpha_i -1)/\alpha_i} G_N^{1/\alpha_i}\delta P^{2/\alpha_i} ~,\end{eqnarray} where $\beta_i$ are numerical factors of order unity. In order each term separately to provide a lower bound on position uncertainty, one should require $\alpha_i \leq 2$. On the other hand, it to be possible to switch off quantum mechanics $\hbar \rightarrow 0$, one has to require $\alpha_i \geq 1$. The correction term $\alpha_i = 1$ is unique in that it does not depend on $\hbar$ and therefore survives even when $\hbar \rightarrow 0$. This correction can be derived as a result of gravitational extension of the length scale as viewed from the background Minkowskian space. 

Indeed, assume particle is localized within the region $\delta X$. Even if particle's mean value of momentum is zero, there is quantum fluctuation in momentum according to the Heisenberg uncertainty relation $\delta P \simeq \hbar /\delta X$. Hence, the energy density over the spherical region with linear size $\delta X$ reads as $\rho = 3\sqrt{\delta P^2 +m^2} /4\pi (\delta X)^3$. One then easily estimates the gravitational field within the region $\delta X$ with constant energy density $\rho$, see \cite{LL}, and finds that the distance $\delta X$ is extended because of gravity as

\begin{equation}\label{gravext} \delta X_{physical} \,=\, \int\limits_0^{\delta X}\frac{dr}{\sqrt{1-\frac{8\pi G_N \rho}{3}\,r^2}} ~. \end{equation} Interpreting from the standpoint of the background Minkowskian space, gravity provides a "repulsive" effect on the wave-packet localised within some region. From Eq.\eqref{gravext} one sees that this background Minkowskian space interpretation makes sense as long as $8\pi G_N \rho (\delta X)^2/3 \ll 1 \Rightarrow \delta X \gg 2G_N\sqrt{\delta P^2 +m^2}$, that is, the width of the wave-packet should be much greater then its gravitational radius. In this case from Eq.\eqref{gravext} one finds

\begin{eqnarray}\label{gravextcorr1} \delta X_{physical} \,=\, \int\limits_0^{\delta X}\frac{dr}{\sqrt{1-\frac{8\pi G_N \rho}{3}\,r^2}} \,\approx\, ~~~~~~~~~\nonumber \\  \int\limits_0^{\delta X}dr \left(1 \,+\, \frac{4\pi G_N \rho}{3}\,r^2 \right) \,=\, \delta X \,+\, \frac{G_N\sqrt{\delta P^2 +m^2}}{3} ~. \end{eqnarray} Assuming in Eq.\eqref{gravextcorr1} $\delta P^2 \gg m^2$ the modified uncertainty relation takes the form originally suggested in the framework of string theory \cite{Veneziano:1986zf}  

\begin{equation}\label{gur2} \delta X \delta P  \,\geq \,
  \frac{1}{2} \,+\, \beta l_P^2\delta P^2~,  \end{equation} hereafter we will set $\hbar =1$. For other {\tt Gedankenexperiments} motivating this sort of (modified) position-momentum uncertainty relation see \cite{Gedanken}.

The correction term in Eq.\eqref{gur2} makes sense even when $\delta P \gtrsim G_N^{-1/2}$. In this case it is motivated by the fact that in high center of mass energy scattering, $\sqrt{s} \gtrsim G_N^{-1/2}$, the production of black holes dominates all perturbative processes \cite{Giddings:2001bu, Dvali:2010bf}, thus limiting the ability to probe short distances. (It is important to notice that at high energies, $\sqrt{s} \gg G_N^{-1/2}$, the black hole production is increasingly a long-distance, semi-classical process). When the black-hole formation becomes likely, what one can say about the position of the observed particle is that the particle was somewhere within the region of the order of horizon. (To make this point more precise, we take into account that the black-hole production becomes certain when the impact parameter $\lesssim G_N\sqrt{s} $ \cite{Giddings:2001bu, Dvali:2010bf}). The gravitational radius of the black hole grows as $r_g \simeq G_N \sqrt{s}$ determining therefore high energy behavior of the position uncertainty.

Let us specify that Planck length and Planck energy are defined as $l_P^2 \equiv \hbar G_N \Rightarrow l_P \approx 10^{-33}$cm; $E_P^2 \equiv \hbar / G_N \Rightarrow E_P \approx 10^{19}$GeV.

\vspace{0.4cm}
\section{Implications for QFT }
\label{incmlqft}

\subsection{Modification at the first quantization level }

Usually, relativistic wave equations are constructed by defining of a four-momentum operator $p_{\mu}$, which in the coordinate representation has the form $p_{\mu} = -i\partial_{\mu}$ \cite{Fradkin:1991ci}. Now let us define the modified three-momentum with respect to the three-dimensional generalization of Eq.\eqref{gur2} in which we simply write $\beta l_P^2 \rightarrow \beta$. Such a generalization that preserves translation and rotation invariance and introduces a finite minimum position uncertainty in all three position variables maybe written as \cite{Kempf:1996nk, Kempf:1996fz, Kempf:2000ac}

\begin{eqnarray} && \left[\widehat{X}^i,\,\widehat{P}^j\right] = i\left( \frac{2\beta \widehat{\mathbf{P}}^2}{\sqrt{1+4\beta \widehat{\mathbf{P}}^2} \,-\, 1}\,\delta^{ij} +2\beta \widehat{P}^i\widehat{P}^j  \right)~, \nonumber \\&& \left[\widehat{X}^i,\,\widehat{X}^j\right] = \left[\widehat{P}^i,\,\widehat{P}^j\right]=0~.\label{minlengthqm}  \end{eqnarray} The deformed $\widehat{\mathbf{X}},\,\widehat{\mathbf{P}}$ operators can be represented in terms of the standard $\widehat{\mathbf{x}},\,\widehat{\mathbf{p}}$ operators in the following way 

\begin{equation}\label{deformedopmultid} \widehat{X}^i = \widehat{x}\,^i\,,~~~~\widehat{P}\,^i = \frac{\widehat{p}\,^i}{1-\beta \widehat{\mathbf{p}}^2}~.  \end{equation} Its Hilbert space realization in the standard-momentum, $\mathbf{p}$, representation has the form

\begin{equation}\label{standardprepresentation} \widehat{X}^i\psi(\mathbf{p}) = i\partial_{p_i} \psi(\mathbf{p})\,,~~~~ \widehat{P}^i\psi(\mathbf{p}) = \frac{p^i}{1-\beta \mathbf{p}^2}\,\psi(\mathbf{p})~, \nonumber \end{equation}  with the scalar product 

\begin{equation}\label{scalarproduct} \langle \psi_1 | \psi_2 \rangle = \int\limits_{\mathbf{p}^2 < \beta^{-1}}d^3p \,\psi^*_1(\mathbf{p})\psi_2(\mathbf{p})~. \nonumber \end{equation}

\noindent The modified field theory takes the form

\begin{widetext} 
\begin{eqnarray}\label{scactionone}  \mathcal{A}[\varPhi] = - \int d^4x \, \frac{1}{2} \left[\varPhi\partial_t^2\varPhi  + \varPhi\widehat{\mathbf P}^2\varPhi 
+ m^2\varPhi^2 \right] \,=\, - \int d^4x \, \frac{1}{2} \left[\varPhi\partial_t^2\varPhi  + \varPhi\frac{-\Delta}{(1+\beta\Delta)^2}\varPhi 
+ m^2\varPhi^2 \right] \,=\nonumber \\ - \int d^4x \, \frac{1}{2} \left[\varPhi\partial_t^2\varPhi  - \varPhi\Delta \sum\limits_{n=0}^{\infty}\left(n + 1 \right)\left(-\beta\Delta\right)^n \varPhi 
+ m^2\varPhi^2 \right] ~.~~~~~~~~~~~~~~~~ \end{eqnarray}\end{widetext}

\noindent The equation of motion reads 

\begin{eqnarray}\label{eqofmot}&&
\left(\partial_t^2 \,+\, \widehat{\mathbf P}^2 \,+\, m^2 \right)\varPhi \,=\, \nonumber \\&&~~~~~~~~ \left(\partial_t^2 \,-\, \frac{\Delta}{(1+\beta\Delta)^2} \,+\, m^2 \right)\varPhi \,=\,0~.
\end{eqnarray}

\noindent  Expressing $\mathbf{p}$ in terms of $\mathbf{P}$  

\[ P^i \,=\, \frac{p^i}{1-\beta \mathbf{p}^2}~,~~\Rightarrow ~~ p^i \,=\, P^i\,\frac{\sqrt{1 \,+\, 4\beta \mathbf{P}^2} \,-\,1}{2\beta \mathbf{P}^2} ~,\] one finds

\begin{equation}   e^{i\mathbf{p}\mathbf{x}} ~~\rightarrow ~~ \exp\left(i\mathbf{x}\mathbf{P}\,\frac{\sqrt{1 \,+\, 4\beta \mathbf{P}^2} \,-\,1}{2\beta \mathbf{P}^2} \right) ~.\label{mlwave}\nonumber \end{equation}

\noindent So, the wave-length is represented now as 

\begin{equation}\label{wavelength} \lambda \,=\, \frac{2\pi}{P} \, \frac{2\beta P^2}{\sqrt{1 \,+\, 4\beta P^2} \,-\,1} ~, \nonumber \end{equation} and is bounded from below $\lambda \geq 2\pi \sqrt{\beta}$ no matter what the value of $P$ is. Thus, 

\begin{eqnarray} \widehat{\mathbf{P}}  \, \exp\left(i\mathbf{x}\mathbf{P}\,\frac{\sqrt{1 \,+\, 4\beta \mathbf{P}^2} \,-\,1}{2\beta \mathbf{P}^2} \right)  \,=\, ~~~~~~~~~~~~~~~~~~~~~ \nonumber \\ P\, \exp\left(i\mathbf{x}\mathbf{P}\,\frac{\sqrt{1 \,+\, 4\beta \mathbf{P}^2} \,-\,1}{2\beta \mathbf{P}^2} \right)~, \nonumber \end{eqnarray} and the solution of Eq.\eqref{eqofmot} is given by \[ \exp\left(i t\sqrt{\mathbf{P}^2 +m^2} - i \mathbf{x}\mathbf{P}\,\frac{\sqrt{1 \,+\, 4\beta \mathbf{P}^2} \,-\,1}{2\beta \mathbf{P}^2} \right) ~. \] Going to the momentum $\mathbf{p}$, the field operator takes the form

\begin{widetext}
\begin{equation} \widehat{\varPhi}(t,\,\mathbf{x}) \,=\, \int\limits_{\mathbf{p}^2 < \beta^{-1}}\, \frac{d^3p}{\sqrt{ (2\pi)^{3}2 \varepsilon_\mathbf{p}}} \, \left[e^{i(\mathbf{p}\mathbf{x} - \varepsilon_\mathbf{p}t)}\,\widehat{a}(\mathbf{p}) \,+\, e^{-i(\mathbf{p}\mathbf{x} - \varepsilon_\mathbf{p}t)}\,\widehat{a}^+(\mathbf{p}) \right] ~,\label{fieldoperator}\end{equation}\end{widetext} where 

\[\varepsilon_\mathbf{p} \,=\, \sqrt{ \frac{\mathbf{p}^2}{(1-\beta \mathbf{p}^2)^2}  +m^2 }~.\]

\noindent So we arrive at the modified dispersion relation 

\begin{eqnarray}\label{dispersionmodrel} \varepsilon^2 &=& \mathbf{P}^2 +m^2 \,=\, \frac{\mathbf{p}^2}{(1-\beta \mathbf{p}^2)^2}  +m^2 \,=\, \nonumber \\&& \sum\limits_{n =0}^{\infty} (1+n)\beta^n \mathbf{p}^{2(n+1)} +m^2~. \end{eqnarray}

\noindent  From Eqs.(\ref{scactionone}, \ref{fieldoperator}) one gets the Hamiltonian

\begin{eqnarray}\label{Hamiltonian}&& H =   \frac{1}{2} \int d^3x \, \left[\varPi^2  + \varPhi\widehat{\mathbf P}^2\varPhi 
+ m^2\varPhi^2 \right]  \,\Rightarrow\,  \widehat{H} \,=  \nonumber \\&&  \int\limits_{\mathbf{p}^2 < \beta^{-1}}d^3p \, \frac{  \varepsilon_\mathbf{p}}{2} \left[\widehat{a}^+(\mathbf{p})\widehat{a}(\mathbf{p}) \,+\, \widehat{a}(\mathbf{p})\widehat{a}^+(\mathbf{p})\right] ~.~~~~~~ \end{eqnarray}

\noindent  The commutation relation $\left[\widehat{\varPi}(t,\,\mathbf{x}), \, \widehat{\varPhi}(t,\,\mathbf{y})\right]$, where as in the standard case $\widehat{\varPi}(t,\,\mathbf{x}) = \dot{\widehat{\varPhi}}(t,\,\mathbf{x})$, takes the form (see Appendix)

\begin{eqnarray}\label{fieldmodiiedcommutator} \left[\widehat{\varPi}(\mathbf{x}), \, \widehat{\varPhi}(\mathbf{y})\right] \,=\, -i\,\widehat{I} \int\limits_{\mathbf{p}^2 < \beta^{-1}}d^3p \,\, \frac{e^{i\mathbf{p}(\mathbf{x} - \mathbf{y})}}{ (2\pi)^{3}} \, = \,  \nonumber \\  \frac{-i\,\widehat{I}}{2\pi^2 r_{xy}} \left[  \frac{1}{r_{xy}^2}\,\sin\left(\frac{r_{xy}}{\sqrt{\beta}}\right) \,-\, \frac{1}{r_{xy}\sqrt{\beta}} \, \cos\left(\frac{r_{xy}}{\sqrt{\beta}}\right) \right]  ~, \end{eqnarray} where $r_{xy} = |\mathbf{x} - \mathbf{y}|$. The Eq.\eqref{fieldmodiiedcommutator} is the "smoothed out" delta function; so in the limit $\beta \rightarrow 0$ one recovers the standard quantization condition. Otherwise speaking, now the operators $\widehat{\varPi}(\mathbf{x}), \, \widehat{\varPhi}(\mathbf{y})$ are non-commuting within the region $|\mathbf{x} - \mathbf{y}|\lesssim \sqrt{\beta}$. Let us notice that quantum field theories with a minimal length scale have also been studied in \cite{Hossenfelder:2003jz}, and they were shown in \cite{Hossenfelder:2007fy} to yield a modification of the equal time commutation relation very similar to the one found here.

Finally let us notice that such a deformation admits a superluminal motion. Namely taking $m=0$, one finds for group velocity (see Eq.\eqref{dispersionmodrel})   

\begin{eqnarray}\label{velocity} \frac{d\varepsilon}{dp} \,=\, \frac{1 \,+\, \beta p^2}{\left(1 \,-\, \beta p^2 \right)^2} \,>\, 1~. \end{eqnarray}

It is worth noticing that the vacuum polarization effect in QED in presence of the classical gravitational background also indicates the superluminal propagation of light \cite{Drummond:1979pp, Khriplovich:1994qj}; for a comprehensive review of this phenomenon see \cite{Shore:2003zc}. (The physical meaning of this effect can be understood in the following way. For in terms of the Feynman diagrams the vacuum polarization effect can be read off as the process of photon decay into virtual electron-positron pares and their subsequent annihilation, the photon acquires an effective size characterized with the Compton wave-length of the electron. It then follows that the photon propagation is affected by gravity when the scale of space-time curvature becomes comparable to the Compton wave-length of electron).

\subsection{Modification at the second quantization level }

Concerning the deformed quantum mechanics, an important issue is to study how it affects the second quantization picture. One the one hand, doing so is necessary to have a self-consistent picture for theories with modified dispersion relations. On the other hand, performing the field quantization with respect to the deformed prescription may result in the long-ranged corrections that can be used for comparing the results with the effective field theory approach. In fact that may be the only way to pick out more realistic models with deformed dispersion relations.

We put aside the modifications at the first quantization level and pose the question how the deformed quantization can be extended to the second quantization level. Field theory is usually understood as a system having non-denumerably many degrees of freedom, which can be represented as a limit of a discretized theory with denumerable degrees of freedom. But it is worth noticing that different discretizations may lead to distinct results when deformed quantization is applied. If the field theory discretization is done through the space discretization then in the limit of a continuous space the effect of deformed quantization simply vanishes \cite{Mania:2009dy}.

Namely, looking at field theory as a limit of a discrete system 

\[ L = \int d^3 x \mathcal{L}(\varPhi,\, \dot{\varPhi},\, \boldsymbol{\bigtriangledown}\varPhi ) \sim \sum\limits_{\mathbf{x}_i} (\delta x)^3 \mathcal{L}(\varPhi_i,\, \dot{\varPhi}_i,\, [\boldsymbol{\bigtriangledown}\varPhi]_i )~,\] (spatial integral is represented as a Riemann-sum over the discrete set of points), one finds \[ P_i = \frac{\partial L}{\partial \dot{\varPhi}_i} = \varPi(\mathbf{x}_i)(\delta x)^3~,\] and the Eq.(\ref{minlengthqm}) to the first order in $\beta$  

\[ [\widehat{\varPhi}_i,\,\widehat{P}_j] = i\left( \delta_{ij}\widehat{I} + \beta \widehat{\mathbf{P}}^2\delta_{ij} + 2\beta \widehat{P}_i \widehat{P}_j \right)~, \] 
takes the form 

\begin{widetext}
\[ \left[\widehat{\varPhi}_i,\,\widehat{\varPi}(\mathbf{x}_j)\right] = i\left( \frac{\delta_{ij}}{(\delta x)^3} \,\widehat{I} + \beta (\delta x)^3  \frac{\delta_{ij}}{(\delta x)^3} \sum\limits_{k} (\delta x)^3 \widehat{\varPi}^2(\mathbf{x}_k) + 2\beta (\delta x)^3 \widehat{\varPi}(\mathbf{x}_i)\widehat{\varPi}(\mathbf{x}_j) \right) ~.\] Taking the limit $(\delta x)^3 \rightarrow 0$ one finds that the corrections to the commutator disappear   

\begin{equation}\label{referisshenishvna} \left[\widehat{\varPhi}(\mathbf{x}),\,\widehat{\varPi}(\mathbf{x}')\right] = i\left( \delta(\mathbf{x} - \mathbf{x}') \,\widehat{I} + \lim\limits_{(\delta x)^3\rightarrow 0} \left[\beta (\delta x)^3   \delta(\mathbf{x} - \mathbf{x}') \int d^3y \,\widehat{\varPi}^2(\mathbf{y}) + 2\beta (\delta x)^3 \widehat{\varPi}(\mathbf{x})\widehat{\varPi}(\mathbf{x}') \right]\right) = i\delta(\mathbf{x} - \mathbf{x}') ~.\end{equation} Instead, one may find it reasonable to replace in Eq.\eqref{referisshenishvna} the limit $(\delta x)^3 \rightarrow 0$ by the $(\delta x)^3 \rightarrow \beta^{3/2}$ that gives 

\begin{equation}\label{refertoinconclusions} \left[\widehat{\varPhi}(\mathbf{x}),\,\widehat{\varPi}(\mathbf{x}')\right] = i\left( \delta(\mathbf{x} - \mathbf{x}')\,\widehat{I} \,+\, \beta^{5/2}  \delta(\mathbf{x} - \mathbf{x}') \int d^3y \,\widehat{\varPi}^2(\mathbf{y}) \,+\, 2\beta^{5/2}  \widehat{\varPi}(\mathbf{x})\widehat{\Pi}(\mathbf{x}') \right)~. \end{equation}

\end{widetext}

\noindent To avoid this sort of ambiguity, we find it more natural to use the oscillator expansion of the field theory for applying of minimum-length deformed formalism at the second quantization level \cite{Mania:2009dy, Berger:2010pj}. Oscillator expansion implies the expansion of the field into plane waves representing free particle wave functions in the Minkowskian background - thus paving the way for defining of quantum (particle) in terms of quantized field.  

After introducing of ladder operators, the Hamiltonian of a neutral scalar field  $\varPhi$ enclosed in a finite volume $l^3$

\[ H = \int\limits_{l^3} d^3x \, \frac{1}{2} \left[ \varPi^2 + \partial_{\mathbf{x}} \varPhi\partial_{\mathbf{x}} \varPhi + m^2  \varPhi^2 \right]~, \] where $\varPi = \dot{\varPhi}$, reduces to 

\begin{equation} \widehat{H} = \frac{1}{2}   \sum\limits_{\mathbf{p}_n}  \varepsilon_{\mathbf{p}_n} \left[ \widehat{a}^+(\mathbf{p}_n)\widehat{a}(\mathbf{p}_n) + \widehat{a}(\mathbf{p}_n)\widehat{a}^+(\mathbf{p}_n) \right] ~, \label{hamintermsaadagger} \nonumber \end{equation} where $\varepsilon_{\mathbf{p}_n} = \sqrt{\mathbf{p}_n^2 + m^2}$. Introducing real variables 

\begin{eqnarray} && \widehat{q}_{\mathbf{p}_n} =\, \sqrt{\frac{l_{\star}}{2\varepsilon_{\mathbf{p}_n} }} \left[ \widehat{a}(\mathbf{p}_n) + \widehat{a}^+(\mathbf{p}_n)\right]~,\nonumber \\ && \widehat{p}_{\mathbf{p}_n} = \, i\sqrt{\frac{\varepsilon_{\mathbf{p}_n}}{2 l_{\star}}} \left[\widehat{a}^+(\mathbf{p}_n) - \widehat{a}(\mathbf{p}_n)\right] ~,\nonumber \end{eqnarray} the Hamiltonian splits into a sum of independent one-dimensional oscillators

\begin{equation} \widehat{H} =   \sum\limits_{\mathbf{p}_n} \left( \frac{l_{\star} \widehat{p}_{\mathbf{p}_n}^{\,2}}{2} + \frac{\varepsilon_{\mathbf{p}_n}^2 \widehat{q}_{\mathbf{p}_n}^{\,2}}{2l_{\star}} \right)~.\label{oscillsum}\end{equation}

The length scale $l_{\star}$ in Eq.\eqref{oscillsum} is of no importance in the framework of standard quantization for the energy spectrum of harmonic oscillator does not depend on its mass, but it appears explicitly in the energy spectrum when the minimum length deformed quantization is applied \cite{Kempf:1996fz}. Namely, for each oscillator in Eq.\eqref{oscillsum} now we have a one-dimensional minimum-length deformed quantization condition  
\begin{equation} \left[\widehat{Q}_{\mathbf{p}_n},\,\widehat{P}_{\mathbf{p}_m}\right] = i\delta_{\mathbf{p}_n\mathbf{p}_m} \left(\widehat{I} \,+\, \beta \widehat{P}^{\,2}_{\mathbf{p}_n}\right)~,\label{mlqm}\nonumber \end{equation} that to the first order in $\beta$ admits the following representation in terms of the standard $\widehat{q}_{\mathbf{p}_n},\, \widehat{p}_{\mathbf{p}_n} $ operators 

\begin{equation} \widehat{Q}_{\mathbf{p}_n} \,=\, \widehat{q}_{\mathbf{p}_n} ~,~~ \widehat{P}_{\mathbf{p}_n} \,=\, \widehat{p}_{\mathbf{p}_n} \left[ \widehat{I} \,+\, \frac{\beta}{3}\,\widehat{p}_{\mathbf{p}_n}^{\,2} \right] ~.\label{onedimsoln}\nonumber \end{equation}

Now replacing $\widehat{p}_{\mathbf{p}_n} \rightarrow \widehat{P}_{\mathbf{p}_n},\, \widehat{q}_{\mathbf{p}_n} \rightarrow \widehat{Q}_{\mathbf{p}_n}$, the Hamiltonian \eqref{oscillsum} gets modified as

\begin{eqnarray}&& \widehat{H} =   \sum\limits_{\mathbf{p}_n} \left( \frac{l_{\star} \widehat{p}_{\mathbf{p}_n}^{\,2}}{2} \,+\, \frac{\varepsilon_{\mathbf{p}_n}^2 \widehat{q}_{\mathbf{p}_n}^{\,2}}{2l_{\star}} \,+\,  \frac{\beta l_{\star} \widehat{p}_{\mathbf{p}_n}^{\,4}}{3}  \right) \,=\, \nonumber \\   && \sum\limits_{\mathbf{p}_n} \left\{  \frac{\varepsilon_{\mathbf{p}_n}}{2} \left[ \widehat{a}^+(\mathbf{p}_n)\widehat{a}(\mathbf{p}_n) + \widehat{a}(\mathbf{p}_n)\widehat{a}^+(\mathbf{p}_n)  \right] \,+\, \right. \nonumber \\&& ~~~~~~~~~~~~~~~~~\left. 
\frac{\beta \varepsilon_{\mathbf{p}_n}^2\left[\widehat{a}^+(\mathbf{p}_n) - \widehat{a}(\mathbf{p}_n)\right]^4}{6l_{\star}} \right\}   ~.\label{oscillsumcorrected}\end{eqnarray}

So, we see that the second quantization with respect to the minimum-length deformed prescription necessarily involves some characteristic length (energy) scale $l_{\star}$ just in the vein of an effective QFT. For the purpose of identifying the length scale $l_{\star},$ one may keep in mind that in view of Eq.\eqref{gur2} the deviation from the standard quantization becomes appreciable at high energies. Therefore it naturally suggests the identification of $l_{\star}^{-1}$ with the characteristic energy scale of the problem under consideration (for more details see \cite{Berger:2010pj}). 

Using the corrected Hamiltonian \eqref{oscillsumcorrected} one may solve Heisenberg equation for the field operator to the first order in $\beta$ and see how the wave-function of the free particle $\langle 0 |\widehat{\varPhi}(t,\,\mathbf{x}) |\mathbf{p} \rangle$ gets modified. This way one arrives at the modified dispersion relation \cite{Berger:2010pj}

\begin{equation}  \varepsilon = \sqrt{\mathbf{p}^2 + m^2} + \beta \, \frac{\mathbf{p}^2 + m^2}{l_{\star}}~.\label{moddisprelsecquant}\end{equation}

\section{Corrections to the Newtonian potential }

Solving the Eqs.\eqref{minlengthqm} to the leading order in $\beta$ for $\widehat{\mathbf{X}},\, \widehat{\mathbf{P}}$ operators, one can not catch the cut-off on $p$. Namely, it is exact solution \eqref{deformedopmultid} that indicates this cut-off. Let us first consider the former case. To the leading order in $\beta$, the Eq.\eqref{dispersionmodrel} implies the modified dispersion relation

\begin{equation}\label{mdrfirst} \varepsilon^2 \,=\, {\mathbf p}^2 \,+\, m^2 \,+\, 2\beta  {\mathbf p}^4 ~, \end{equation} that for the gravitational potential gives \cite{Pais:1950za}

\begin{eqnarray}\label{mpfirst}&& \int \frac{d^3k}{(2\pi)^3} \,\frac{4\pi \, e^{i\mathbf{k}\mathbf{x}}}{\mathbf{k}^2 \,+\, 2{\beta} {\mathbf k}^4 } \,=\,  \nonumber\\&& \int \frac{d^3k}{(2\pi)^3} \,\left( \frac{4\pi \, e^{i\mathbf{k}\mathbf{x}}}{\mathbf{k}^2 } - \frac{4\pi \, e^{i\mathbf{k}\mathbf{x}}}{\mathbf{k}^2 + 1/2{\beta}} \right) \,=\,\nonumber\\&& \frac{1}{r} \,-\, \frac{e^{-r/\sqrt{2\beta}}}{r} ~.\end{eqnarray} So, the correction due to Eq.(\ref{mdrfirst}) appears to be a very short-ranged repulsive Yukawa potential. One could obtain this result by solving the Poisson's equation for minimum-length modified electromagnetic field \cite{Tkachuk:2007zz}.

Now, by using the cut-off $\mathbf{k}^2 < \beta^{-1}$ and the exact dispersion relation Eq.\eqref{dispersionmodrel}, one gets the following expression (see Appendix)

\begin{figure}[h]
\includegraphics[width= 8cm, height= 7cm]{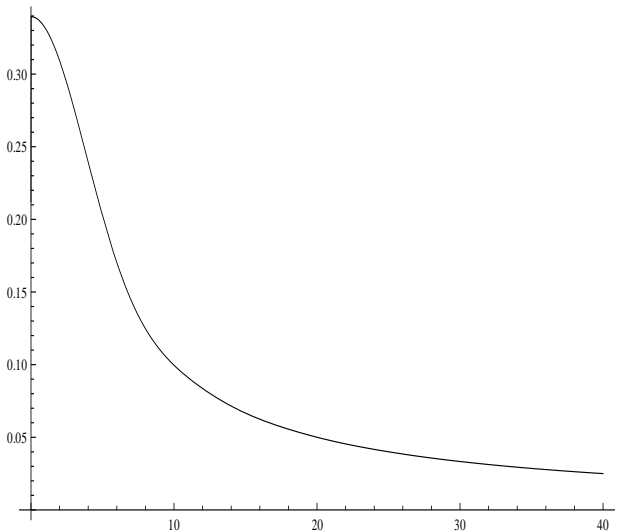}\\
\caption{Vertical axis: $V(r)\sqrt{\beta}$ , horizontal axis: $x = r/\sqrt{\beta}$.}
\label{potential}
\end{figure}

\begin{widetext}
\begin{eqnarray}\label{shmpotcut} V(r) \,=\,  \int\limits_{\mathbf{k}^2 < \beta^{-1}} \frac{d^3k}{(2\pi)^3} \,4\pi \, e^{i\mathbf{k}\mathbf{x}} \left( \frac{1}{\mathbf{k}^2} \,-\, 2\beta \,+\, \beta^2\mathbf{k}^2  \right)  \,=\, ~~~~~~~~~~~~~~~~~~~~~~~~~~~~~~~~~~~~~~~~~~~~~~~~~~~~~~~~ \nonumber \\ \frac{2}{\pi r} \left[  \text{Si}\left(\frac{r}{\sqrt{\beta}}\right)  \,+\, \frac{2\sqrt{\beta}}{r} \, \cos\left(\frac{r}{\sqrt{\beta}}\right)  \,-\, \frac{2\beta}{r^2}\,\sin\left(\frac{r}{\sqrt{\beta}}\right) \,-\, \frac{\beta^{3/2}\left(r^2/\beta -6\right) }{r^3}\, \cos\left(\frac{r}{\sqrt{\beta}} \right) \,+\, \frac{3 \beta^2 \left(r^2/\beta -2 \right) }{r^4}\,\sin\left(\frac{r}{\sqrt{\beta}} \right) \right] ~.~~\end{eqnarray}\end{widetext} When $x \rightarrow 0$ the sin-integral function behaves as $\text{Si}(x)  \sim x$. Using this asymptotic expression one finds that for $r \ll \sqrt{\beta}$ the Eq.\eqref{shmpotcut} behaves as 

\begin{eqnarray}\label{shortasympot}  V\left(r \ll \sqrt{\beta}\right)\, = \, \frac{1.2}{\pi\sqrt{\beta}} \,+\, O\left(r^2\right) ~. \end{eqnarray} So, the gravitational force vanishes when $r \rightarrow 0$. The behaviour of $V(r)$ is shown in Fig. \ref{potential}.

The other modification to the Newtonian potential arises from Eq.\eqref{moddisprelsecquant}, where $l_{\star}$ denotes an intrinsic (IR) length scale to the problem under consideration. How it can be applied to a concrete problem? Certainly a subtle point is to identify the scale $l_{\star}$. Using the modified propagator with respect to the Eq.\eqref{moddisprelsecquant} one finds   

\begin{eqnarray}\label{mpsecond}&& \int \frac{d^3k}{(2\pi)^3} \,\frac{4\pi \, e^{i\mathbf{k}\mathbf{x}}}{k^2 \,+\, 2{\beta}  k^3 / l_{\star} } = \nonumber\\&&  \int \frac{d^3k}{(2\pi)^3}  \,4\pi \, e^{i\mathbf{k}\mathbf{x}} \left( \frac{1}{k^2} \,-\, \frac{2{\beta}}{kl_{\star}}  \,+\, O\left(\beta^2 \right)  \right) \,=\, \nonumber\\ && \frac{1}{r} \,-\, \frac{4\beta }{\pi r^2 l_{\star}} \,+\, O\left(\beta^2 \right)  ~.\end{eqnarray} For this particular problem there is only one naturally appearing length scale, it is $r$. So one can set unambiguously $l_{\star} \simeq r$. With this identification one finds again a repulsive correction to the Newtonian potential of the form $\sim \beta/r^3$. Such a correction to the Newtonian potential arises at the one loop level in an effective field theory approach to general relativity \cite{BjerrumBohr:2002kt, Park:2010pj}, but in this approach the question of the sign for this correction remains controversial.

\section{Implications for black holes} 

\subsection{Black hole remnants and regularization of gravitational singularities}

One might feel unsatisfied with the explanation of black hole emission ceasing due to Eq.\eqref{gur2} proposed in paper \cite{Adler:2001vs}, that the temperature for the black hole emission estimated via the relation \eqref{gur2} becomes a complex quantity when an evaporating black hole mass approaches Planck scale.

Instead, we notice that the repulsive short ranged gravitational correction may be responsible for black hole radiation halt as, on the one hand, it implies the vanishing of gravitational radius, $r_g$, when black hole evaporates down to the Planck mass and, on the other hand, the surface gravity that determines Hawking temperature becomes finite for $r_g =0$ (or even zero). Thus, the appearance of a short ranged repulsive correction to the Newtonian potential may be a proper understanding of the reason why Eq.\eqref{gur2} may imply black hole remnants.

Let us first consider the modified Schwarzschild space-time with respect to the potential \eqref{mpfirst} (for definiteness in this paragraph we set $\beta = l_P^2/2$)

\begin{eqnarray}\label{modschwarzschild} ds^2 \,=\, f(r)dt^2 \,-\,f(r)^{-1}dr^2 \,-\, r^2 d\Omega^2 ~,\nonumber  \end{eqnarray} where \begin{eqnarray} f(r) \,=\, 1 \,-\,\frac{2l_P^2 m}{r} \,+\, \frac{2l_P^2m \, e^{-r/l_P}}{r} ~. \nonumber \end{eqnarray} The equation for the gravitational radius takes the form 

\begin{equation}\label{modhorizon} r \,-\, 2l_P^2 m  \,=\, -2l_P^2 m \, e^{-r/l_P} ~. \end{equation}

\noindent On the left hand side there is a straight line $w = r \,-\, 2l_P^2 m$ passing through the points $(0, -2l_P^2m)$ and $(2l_P^2m, 0)$, and on the right there is a monotonically increasing function $w= -2l_P^2 m \, e^{-r/l_P}$ that starts from the same point $(0, -2l_P^2m)$ and asymptotically approaches real axes. The derivative of the latter is a positive monotonically decreasing function. So, if the derivative of $w= -2l_P^2 m \, e^{-r/l_P}$ at the initial point $(0, -2l_P^2m)$ is smaller than the slope of $w = r \,-\, 2l_P^2 m$, that is unity, then these two curves never intersect each other and therefore Eq.\eqref{modhorizon} will have the only solution $r=0$. The derivative of $w= -2l_P^2 m \, e^{-r/l_P}$ at $r = 0$ is $w'(0) = 2l_Pm$, which is less then unity for $m< E_P/2$. Thus, when the black hole evaporates down to this mass ($m_{remnant} = E_P/2$), the horizon approaches zero (disappears) and there is no Hawking radiation any more. If so, the question arises what is the exit temperature for such a black hole remnant. Using the relation 

\begin{eqnarray}\label{shortdforce}&&  \frac{d}{dr} \left( \frac{1}{r} \,-\, \frac{e^{-r/\sqrt{2\beta}\,l_P}}{r}\right) \,=\,  \nonumber \\&&~~~~ -\,\frac{1}{4\beta l_P^2} \,-\, \sum\limits_{n=3}^{\infty}\,\frac{n-1}{n!} \, \frac{(-\,r)^{n-2}}{\left(\sqrt{2\beta}\,l_P\right)^n}  ~, \nonumber \end{eqnarray} one gets for the exit temperature (that is, Hawking temperature for $r_g=0$)

\begin{equation}\label{Hawking} T_{exit}  \,=\, \left. \frac{f'(r_g(m))}{4\pi}\right|_{r_g =0} \,=\, \frac{m_{remnant}}{4\pi} ~. \nonumber    \end{equation}

\begin{figure}[t]
\includegraphics[width= 8cm, height= 7cm]{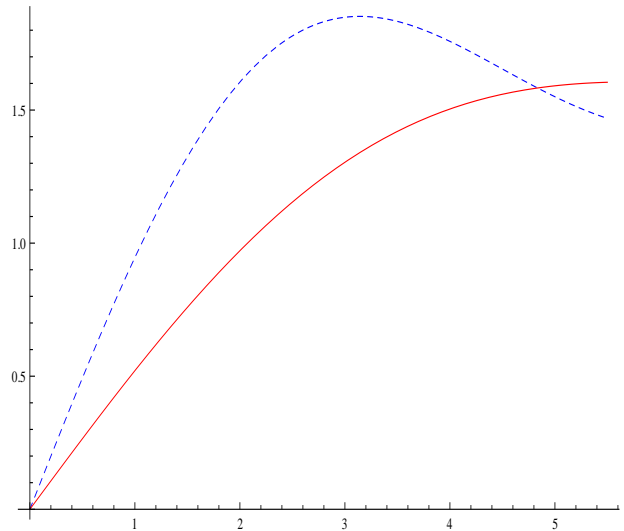}\\
\caption{Dashed blue line - $\text{Si}(x)$, red line - $\mathcal{V}(x)$.   }
\label{figure}
\end{figure}

Now let us consider modified Schwarzschild space-time with respect to Eq.\eqref{shmpotcut}: $f(r) = 1 - 2l_P^2mV(r)$. Denoting in Eq.\eqref{shmpotcut} the expression in square brackets with $\mathcal{V}(x)$, where $x$ stands for $x=r/\sqrt{\beta}$, the equation for the gravitational radius takes the form 

\begin{eqnarray}\label{moreprecisehorizon} \frac{\pi\sqrt{\beta} \, x }{4l_P^2 m} \,= \, \mathcal{V}(x)  ~.\end{eqnarray} Looking at the behaviour of function $\mathcal{V}(x)$, see Fig. \ref{figure}, it is red line in the figure while dashed blue line represents $\text{Si}(x)$ alone, one concludes in a similar way that when black hole evaporates to the Planck mass, the horizon disappears. Namely, the derivative of the expression on the right-hand side in Eq.\eqref{moreprecisehorizon} at $x=0$ is $0.6$ (see Eq.\eqref{shortasympot}). Looking at the figure, one infers that the black hole horizon disappears when $m < E_P \pi \sqrt{\beta} /2.4 l_P$. Then from Eq.\eqref{shortasympot} one infers that the exit temperature for such a black hole remnant is zero. (Looking at the behaviour of $\text{Si}(x)$ in Fig. \ref{figure} and observing that $\text{Si}(x) = x - x^3/18 + O\left(x^5\right)$ one infers that similar results about the black hole remnants will be obtained if only $\text{Si}(x)$ is kept in Eq.\eqref{shmpotcut}).

\subsection{Corrections to the black hole entropy  }

It is easy to see that the increment of energy given by Eq.\eqref{moddisprelsecquant} results in the logarithmic correction to the black hole entropy. By taking into account that $\varepsilon \propto T$ and the only appropriate length scale at hand is $l_{\star} \sim  T^{-1}$, from Eq.\eqref{moddisprelsecquant} one infers that black hole emission temperature gets increased $T \rightarrow T + \beta l_P^2 T^3$. Hence, to the first order in $l_P^2$ the entropy $dS = dM/T \rightarrow dM/T - \beta l_P^2 T dM$ acquires a logarithmic correction 

\[ S = \pi\left(\frac{r_g}{l_P}\right)^2  - \gamma  \ln\left(\frac{r_g}{l_P}\right)~,\] where $\gamma$ is a (positive) number of order unity.

Another peace of correction to the black hole entropy arises from the fact that the minimum length deformed quantization implies a backscattering effect for neutral particles \cite{Berger:2010pj}. Because of this, there is a backscattered part of the evaporated mass, $dM \propto - E_P^2 dT /T^2$, of the order of $dM_{+} \propto T^2 dT /E_P^2$ that implies the entropy correction of the form $\sim l_P^2/r_g^2$. 

Combining together, we get a well known entropy expression

\begin{equation}\label{entropycorrections} S = \pi\left(\frac{r_g}{l_P}\right)^2  - \gamma  \ln\left(\frac{r_g}{l_P}\right) + \eta \left(\frac{l_P}{r_g}\right)^2 + \mbox{const.} ~,\end{equation} ($\gamma$ and $\eta$ are numerical factors of order unity) obtained in loop quantum gravity \cite{lqgbhent} and in a tunnelling formalism approach to the black hole emission \cite{tunbhent}. It should be noted that backscattering effect does not affect the thermal character of the radiation, for more detail see \cite{Berger:2010pj}.

\section{Modified dispersion relations vs. the maximally localised states }

We saw that that both the above considered modified dispersion relation as well as cut-off result in the modified Schwarzschild solution that for the particle with the mass smaller than the Planck energy does not posses event horizon. So, such solution covers the whole region with respect to $r$. Looking at the Poisson's equation for the gravitational potential  

\begin{eqnarray}\label{grpoisson} \Delta V(r) \,=\, -\,4\pi l_P^2 \rho(r) ~,   \end{eqnarray}

\noindent  one easily concludes that for the Eq.\eqref{shortasympot} the distribution function for particle is not any more $\delta(\mathbf{r})$ but some smeared-out version of it. 

Noticing that 
\begin{eqnarray} &&\int\limits_{\mathbf{k}^2 < \beta^{-1}} \frac{d^3k}{(2\pi)^3} \,4\pi \, \frac{e^{i\mathbf{k}\mathbf{r}}}{\mathbf{k}^2} \,=\, \frac{2 \, \text{Si}\left(r/\sqrt{\beta}\right)}{\pi r}  ~, \nonumber  \\&& -2\beta \int\limits_{\mathbf{k}^2 < \beta^{-1}} \frac{d^3k}{(2\pi)^3} \,4\pi \, e^{i\mathbf{k}\mathbf{r}}  \,=\, \nonumber \\&& ~~~~~~~~~~~~~~~~~~~~~~~~ \beta \, \frac{\partial}{\partial \mathbf{r}} \frac{\partial}{\partial \mathbf{r}}  \int\limits_{\mathbf{k}^2 < \beta^{-1}} \frac{d^3k}{(2\pi)^3} \,4\pi \, \frac{e^{i\mathbf{k}\mathbf{r}}}{\mathbf{k}^2}~, \nonumber \\&& \beta^2 \int\limits_{\mathbf{k}^2 < \beta^{-1}} \frac{d^3k}{(2\pi)^3} \,4\pi \, e^{i\mathbf{k}\mathbf{r}} \, \mathbf{k}^2   \,=\, \nonumber \\&&~~~~~~~~~~~~~~~ \beta^2 \, \frac{\partial}{\partial \mathbf{r}} \frac{\partial}{\partial \mathbf{r}}\frac{\partial}{\partial \mathbf{r}}\frac{\partial}{\partial \mathbf{r}}  \int\limits_{\mathbf{k}^2 < \beta^{-1}} \frac{d^3k}{(2\pi)^3} \,4\pi \, \frac{e^{i\mathbf{k}\mathbf{r}}}{\mathbf{k}^2}~, \nonumber \end{eqnarray}

\noindent  the potential can be written as
\begin{eqnarray}&& V_{\beta}(r) \,=\, \int\limits_{\mathbf{k}^2 < \beta^{-1}} \frac{d^3k}{(2\pi)^3} \,4\pi \, e^{i\mathbf{k}\mathbf{r}} \left( \frac{1}{\mathbf{k}^2} \,-\, 2\beta \,+\, \beta^2\mathbf{k}^2  \right) \, = \, \nonumber \\&& \frac{2 \, \text{Si}\left(r/\sqrt{\beta}\right)}{\pi r}  \,+\,  \beta \, \frac{\partial}{\partial \mathbf{r}} \frac{\partial}{\partial \mathbf{r}}  \, \frac{2 \, \text{Si}\left(r/\sqrt{\beta}\right)}{\pi r}  \,+\, \nonumber \\&& ~~~~~~~~~~~~~~~~~~~~~~~~~~~~~~~ \beta^2 \, \frac{\partial}{\partial \mathbf{r}} \frac{\partial}{\partial \mathbf{r}}\frac{\partial}{\partial \mathbf{r}}\frac{\partial}{\partial \mathbf{r}}      \, \frac{2 \, \text{Si}\left(r/\sqrt{\beta}\right)}{\pi r}  ~.\nonumber \end{eqnarray} 

\noindent So, the delta function describing the distribution for point-like particle will be replaced by 

\begin{eqnarray}  \delta_{\beta}(\mathbf{r})  \,=\, - \,\frac{1}{4\pi} \left[\Delta \, \frac{2 \, \text{Si}\left(r/\sqrt{\beta}\right)}{\pi r}  \,+\,  \beta \, \Delta^2  \, \frac{2 \, \text{Si}\left(r/\sqrt{\beta}\right)}{\pi r}  \,+\, \right. \nonumber \\ \left. \beta^2 \, \Delta^3      \, \frac{2 \, \text{Si}\left(r/\sqrt{\beta}\right)}{\pi r}  \right] ~. ~\nonumber   \end{eqnarray}

For more precise description for the characteristics of the maximally localized source (energy-momentum tensor) one may use exact general-relativistic equations for the {\tt ansatz} \eqref{modschwarzschild}, see \cite{LL} 

\begin{eqnarray}&& 8\pi l_P^2 T^t_t  \,=\,  8\pi l_P^2 T^r_r \,=\, -f \left(\frac{1}{r^2} \,+\,\frac{f'}{rf} \right) \,+\,\frac{1}{r^2}  ~,\nonumber \\&& 8\pi l_P^2 T^{\theta}_{\theta}  \,=\,  8\pi l_P^2 T^{\varphi}_{\varphi} \,=\, -\,\frac{f}{2}\left(\frac{f''}{f} \,+\, \frac{2f'}{rf}  \right) \nonumber ~, \end{eqnarray}

\noindent  and substitute in it $f(r) \,=\, 1 \,-\, 2l_P^2m V(r)$.

So, there may be two ways for defining the back reaction on gravity due to concept of minimum length. One of them is to implement the concept of minimum length in quantum theory and see how the propagator and the resulting potential will be modified and the other one is to take a maximally localized state, which is some sort of smeared-out delta function and find for it the corresponding gravitational field \cite{Nicolini:2005vd, Ansoldi:2008jw, Spallucci:2011rn}.

\section{Towards the UV finite QFT}

It has been conjectured long ago that the implementation of a minimum length in quantum theory in an appropriate way \cite{Heisenberg:1996bv}, or taking account of the universal and non-linear coupling of gravitation to matter \cite{Deser:1957zz}, may provide a natural mechanism for the damping of UV infinities in QFT. This problem in the framework of minimum length deformed quantum mechanics was addressed in \cite{Kempf:1996nk}. In that paper, from the very outset the Euclidean theory is assumed and the Eq.\eqref{minlengthqm} is extended for all $X^{\mu},\,P^{\mu}$. This approach makes somewhat obscure to understand what sort of modification is implied for the Minkowskian theory. Let us try to address this problem immediately in the Minkowskian case.

For in the operational sense the time is understood by means of the periodic motion, the existence of the minimum length automatically implies the presence of the minimum time. Indeed, it was noticed long ago that there is a limit in the time resolution when synchronizing clocks \cite{Mead:1964zz} (this argument is also reviewed in \cite{Garay:1994en}). In view of this conclusion one might naturally expect the modification of time energy uncertainty relation of the form

\begin{equation}\label{tegur} \delta t \, \delta E  \,\geq \, \frac{1}{2} \,+\, \beta' t_P^2\delta E^2~, \end{equation} where $\beta'$ is a numerical factor of order unity. 

Writing Eq.\eqref{tegur} in the form 

\[ \delta X^0 \delta P_0 \geq \frac{1}{2} \left(1 \,+\, \beta' \delta P_0^2 \right)~,\] where for brevity we have just replaced $\beta't_P^2 \rightarrow \beta'/2$, one can readily find the solution of this algebra in terms of standard $x^0,\,p_0$ operators \cite{Kempf:1996nk} 

\begin{eqnarray}\label{defxopo} && X^0 = x^0\,,~~ P_0 =  \beta'^{-1/2}\tan\left(p_0\sqrt{\beta'}\right) ~.\end{eqnarray} From Eq.\eqref{defxopo} one gets that there is a cut-off: $p_0 < \pi /2\sqrt{\beta'}$. 

Now let us bring into field-theory consideration Eq.\eqref{defxopo}. Following the discussion of Section \ref{incmlqft}, the action functional takes the form \begin{widetext} 
\begin{eqnarray}\label{scaction}  \mathcal{A}[\varPhi] = \int d^4x \, \frac{1}{2} \left[\varPhi P_0^2\varPhi  - \varPhi\widehat{\mathbf P}^2\varPhi 
- m^2\varPhi^2 \right] \,=\,  \int d^4x \, \frac{1}{2} \left[\varPhi \frac{\tan^2\left(i\partial_t\sqrt{\beta'}\right)}{\beta'} \varPhi  + \varPhi\frac{\Delta}{(1+\beta\Delta)^2}\varPhi 
- m^2\varPhi^2 \right] \,=\nonumber \\ - \int d^4x \, \frac{1}{2} \left[\varPhi \left(\partial_t^2  - \frac{2\beta'}{3}\,\partial_t^4  + \frac{7\beta'^2}{45}\,\partial_t^6 + O\left(\beta'^3\right) \right)\varPhi  - \varPhi \left( \Delta -2\beta\Delta^2  + 3\beta^2\Delta^3 + O\left(\beta^3\right)\right) \varPhi 
+ m^2\varPhi^2 \right] ~.~~~~~~~~~~~~~~~~ \end{eqnarray}\end{widetext} The equation of motion now reads 

\begin{eqnarray}  \left( \frac{\tan^2\left(i\partial_t\sqrt{\beta'}\right)}{\beta'}   + \frac{\Delta}{(1+\beta\Delta)^2} 
- m^2 \right)\varPhi \,=\, 0~. \nonumber \end{eqnarray} The solution of this equation has the form

\begin{eqnarray}\label{mod2wavefuncsecond} \exp\left(i t\varepsilon  - i \mathbf{x}\mathbf{P}\,\frac{\sqrt{1 \,+\, 4\beta \mathbf{P}^2} \,-\,1}{2\beta \mathbf{P}^2} \right) ~, \nonumber  \end{eqnarray} where 

\begin{eqnarray}\label{cutoffonlargep}  \tan^2\left(\varepsilon \sqrt{\beta'}\right) \,=\, \beta' \left( \mathbf{P}^2  \,+\, m^2\right) ~. \nonumber  \end{eqnarray}

From Eq.\eqref{mod2wavefuncsecond} one finds the velocity of the massless particle of the form

\begin{eqnarray}\label{veloctwo} \frac{d\varepsilon}{dp} \,=\, \frac{1 \,+\, \beta p^2}{\left(1 \,-\, \beta p^2 \right)^2 \,+\, \beta' p^2} ~. \end{eqnarray} We see that at the expense of $\beta'$ one can avoid the superluminal motion. Namely, by taking $\beta' = 3\beta$ one gets that the velocity of light is always smaller than unity and varies from $1$ to $2/3$ when the momentum $p$ runs over the interval $(0, \beta^{-1/2})$.

Usually the UV divergences in QFT are caused by the singular behaviour of the propagator

\begin{eqnarray} D_c(x_1 - x_2) \,=\,  \int \frac{dp_0  d^3p}{(2\pi)^4} \,   \frac{e^{ip_{\mu}(x_1^{\mu} - x_2^{\mu})}}{p_0^2 -\mathbf{p}^2 -m^2 +i\epsilon}~,\nonumber \end{eqnarray}  on the light-cone: $x_1 - x_2 = 0$, \cite{Bogolyubov:1980nc}. Now in view of the Eq.\eqref{scaction}, the propagator will have the form 

\begin{eqnarray}  \label{propagator} \propto \int\limits_{|p_0|< \beta'^{-1/2}} dp_0 \int\limits_{\mathbf{p}^2< \beta^{-1}}  d^3p  \,\,   \frac{e^{ip_{\mu}(x_1^{\mu} - x_2^{\mu})}}{P_0^2 -\mathbf{P}^2 -m^2 +i\epsilon}~, \end{eqnarray} which is certainly finite for $x_1 - x_2 = 0$.

\section{Concluding remarks}

The concept of minimum length (implying the impossibility of measuring the particle position to a better accuracy than this length scale) naturally suggests the modification of the position-momentum uncertainty relation in such a way as to get minimum position uncertainty from this relation. In a pure phenomenological way one might imagine many such possibilities, Eq.\eqref{gur1}. But in Eq.\eqref{gur1} one of the combinations for the gravitational correction is naturally singled out by the fact that it does not depend on $\hbar$ and thus survives even in the classical limit Eq.\eqref{gur2}. Despite the fact that this sort of modification is supported by many {\tt Gedankenexperiment} considerations, it seems very welcoming to have some low energy phenomena that could be reproduced by this modification. Such "universal" criteria may be the large-distance gravitational corrections obtained in the framework of an effective field theory approach to general relativity. But as it was pointed out above, in the framework of this approach the question of the sign for the correction term remains a subtle point \cite{BjerrumBohr:2002kt, Park:2010pj} see also \cite{Donoghue:1993eb}. As discussed above, the minimum-length deformed quantum mechanics favours the repulsive correction to the Newtonian potential.

In trying to estimate corrections to the Newtonian potential due to minimum-length deformed quantum mechanics, we have assumed the following universality features. First gravity is treated on the same footing as the other interactions, that is, gravitational interaction is assumed to be mediated by graviton, and second, the modified dispersion relation arising from the minimum-length deformed quantum mechanics is assumed to be applicable for all particles, including graviton. While for matter fields we can immediately implement the deformed momentum into field theory (see for instance \cite{Kober:2010sj}), for the gravitational field we still do not know how (beyond the linearized gravity) this kind of implementation can be done.

Short distance behaviour of the corrected Newtonian potential indicates that the horizon of the black hole disappears when it evaporates down to the Planck mass and the corresponding emission temperature becomes zero. That is, the zero-temperature black hole remnants will be left behind the black hole evaporation. Such remnants might be considered as a possible candidates for the dark-matter \cite{Chen:2002tu}. Interestingly enough, the minimum-length quantum mechanics allows to recover both well known corrections for black hole entropy obtained previously in different approaches \cite{lqgbhent, tunbhent}. One more interesting fact regarding the short distance corrections to the Newtonian potential is that now the gravitational force does not diverge at $r=0$ but rather it becomes zero. However, since the minimum-length deformed position-momentum uncertainty relation does not allow to probe the potential at a given point, what can be measured is just the averaged potential over the length scale $\sqrt{\beta}$.

It is important to notice that second quantization with respect to the minimum-length deformed prescription involves the scale $l_{\star}$, which should always be replaced with the characteristic energy/length scale of the problem under consideration. So, one should not have a misconception that $l_{\star}$ as some IR cut-off enters the final expressions. The above considered examples explicitly demonstrate how this prescription works. In estimating the corrections to the potential $l_{\star}$ was identified as distance $r$ while in considering black hole emission it was set by the characteristic energy scale of the problem, $l_{\star} = T^{-1}$, where $T$ is the emission temperature. Similarly, for the corrections to black-body radiation the characteristic energy scale that should be used for identifying $\l_{\star}$ is the temperature of radiation \cite{Mania:2009dy}.

It is worth noticing that if only position-momentum uncertainty relation is deformed the minimum-length deformed quantum mechanics admits a superluminal motion. If we deform the time-energy uncertainty relation too, it becomes possible to avoid the superluminal motion. Moreover, this way one arrives at an UV finite QFT. Should be noticed that the theory given by Eq.\eqref{scaction} is ghost free (usually ghosts result in the additional poles of the propagator). If we truncate the series in this equation at some power of $\beta'$ then we may face the problem of ghosts.

Recently a new paper addressing the question of field quantization with respect to the minimum-length deformed prescription appeared on arXive: \cite{Kober:2011uj}. Author considers various possible minimum-length deformed quantization rules immediately in terms of $\widehat{\varPhi},\,\widehat{\varPi}$ operators but does not consider their possible counterparts in quantum mechanics and vice verse (we mean the conventional transition from QM to QFT that implies considering QFT as a continuous limit of some discretization scheme, see for instance Eq.\eqref{refertoinconclusions} above).

Finally let us mention a couple of papers addressing the question of the corrections to the Newtonian (Coulomb) potential for different dispersion relations \cite{Helling:2007zv, AmelinoCamelia:2010rm}.

For plotting we have used the package Mathematica 8: http://www.wolfram.com/mathematica/.

\vspace*{0.5cm}
\noindent {\bf Acknowledgements:} The author is grateful to Professor Micheal S. Berger (Indiana U.) for many useful discussions. Useful conversations with Professors Iosif B. Khriplovich, Zurab K. Silagadze (Novosibirsk, IYF) and Zurab Berezhiani (L'Aquila U.) are also acknowledged.

\section*{Appendix}

To estimate the integral 

\begin{eqnarray} \int\limits_{\mathbf{k}^2 < \beta^{-1}} \frac{d^3k}{(2\pi)^3} \,4\pi \, e^{i\mathbf{k}\mathbf{r}} \left( \frac{1}{\mathbf{k}^2} \,-\, 2\beta \,+\, \beta^2\mathbf{k}^2  \right)  ~, \nonumber \end{eqnarray} let us introduce a coordinate system with $z$ axes along $\mathbf{r}$. Then introducing spherical coordinates 

\[k_x \,=\, k \sin\theta \cos\varphi~,~~k_y \,=\, k\sin\theta \sin\varphi~,~~ k_z  \,=\, k\cos\theta~,\] 

\noindent we get $\mathbf{k}\mathbf{r} = k z \cos\theta$ and $d^3k = k^2 \sin\theta \,dk\, d\theta \, d\varphi$,  

\begin{eqnarray}&& \int\limits_{\mathbf{k}^2 < \beta^{-1}} \frac{d^3k}{(2\pi)^3} \,4\pi \, e^{i\mathbf{k}\mathbf{r}} \left( \frac{1}{\mathbf{k}^2} \,-\, 2\beta \,+\, \beta^2\mathbf{k}^2  \right) \, = \, \nonumber \\&&  \frac{1}{\pi} \int\limits_0^{\beta^{-1/2}}  dk   \left( 1 \,-\, 2\beta k^2 \,+\, \beta^2 k^4  \right)\int\limits_0^{\pi} d\theta \, \sin\theta  \, e^{ikz\cos\theta} ~.\nonumber \end{eqnarray}

Denoting $\cos\theta = t$ one finds

\begin{eqnarray}&& \int\limits_0^{\pi} d\theta \, \sin\theta  \, e^{ikz\cos\theta} \,=\, \int\limits_{-1}^1 dt e^{ikzt} \,=\, \nonumber \\ &&  \int\limits_{-1}^1 dt \cos(kzt) \,=\, \frac{2\sin(kz)}{kz} ~.\nonumber \end{eqnarray}

Therefore the integral reduces to 

\begin{widetext}
\begin{eqnarray}&& \frac{2}{\pi r} \int\limits_0^{\beta^{-1/2}}  dk   \left( 1 \,-\, 2\beta k^2 \,+\, \beta^2 k^4  \right)\frac{\sin(kr)}{k} \,=\, \nonumber \\ &&  \frac{2}{\pi r} \left[  \text{Si}\left(\frac{r}{\sqrt{\beta}}\right)  \,+\, \frac{2\sqrt{\beta}}{r} \, \cos\left(\frac{r}{\sqrt{\beta}}\right)  \,-\, \frac{2\beta}{r^2}\,\sin\left(\frac{r}{\sqrt{\beta}}\right) \,-\, \frac{\beta^{3/2}\left(r^2/\beta -6\right) }{r^3}\, \cos\left(\frac{r}{\sqrt{\beta}} \right) \,+\, \frac{3 \beta^2 \left(r^2/\beta -2 \right) }{r^4}\,\sin\left(\frac{r}{\sqrt{\beta}} \right) \right]  ~.\nonumber \end{eqnarray}\end{widetext}

\noindent Similarly, one finds

 \begin{eqnarray}&& \int\limits_{\mathbf{k}^2 < \beta^{-1}} \frac{d^3k}{(2\pi)^3} \, e^{i\mathbf{k}\mathbf{r}} \,=\, \frac{1}{2\pi^2 r}  \int\limits_0^{\beta^{-1/2}}  dk \,k \sin(kr)  \, =\, \nonumber \\&& 
\frac{1}{2\pi^2 r} \left[  \frac{1}{r^2}\,\sin\left(\frac{r}{\sqrt{\beta}}\right) \,-\, \frac{1}{r\sqrt{\beta}} \, \cos\left(\frac{r}{\sqrt{\beta}}\right) \right]  ~. \nonumber \end{eqnarray}

It is important to notice that in the limit $\beta \rightarrow 0 $ this expression approaches $\delta(\mathbf{r})$. So, working out the large distance behavior of this expression, $r \gg \sqrt{\beta}$, which is dominated by the Fourier modes with wave-lengths much greater than the $\sqrt{\beta}$ and therefore almost insensitive to the existence of cut-off, one infers that large distance behavior of this expression is damped not simply as $1/r^2$ but much more stronger because of the quickly oscillating $\cos\left(r/\sqrt{\beta}\right)$ and $\sin\left(r/\sqrt{\beta}\right)$. Namely, passing the minimum-length $\sqrt{\beta}$, the $\cos\left(r/\sqrt{\beta}\right)$ and $\sin\left(r/\sqrt{\beta}\right)$ change their sign about $r/\sqrt{\beta}$ times, so averaging over the minimum-length gives simply zero. Thus, for estimating the large distance behavior of the potential we can simply ignore the cut-off

\begin{eqnarray}&& \int \frac{d^3k}{(2\pi)^3} \,4\pi \, e^{i\mathbf{k}\mathbf{r}} \left( \frac{1}{\mathbf{k}^2} \,-\, 2\beta \,+\, \beta^2\mathbf{k}^2  \right) \,=\, \nonumber \\ && \frac{1}{r} \,-\, 8\pi\beta \delta(\mathbf{r})  \,+\, \frac{24 \beta^2}{r^5}  ~, \nonumber \end{eqnarray} that after omitting the $\delta(\mathbf{r})$ term gives the large distance behavior.

\end{document}